# Spatial and temporal scaled physical modeling of fluid convection using hypergravity

Jinlong Li[1], Wenjie Xu[1]*, Yunmin Chen[1]*, Liangtong Zhan[1], Yingtao Hu[1], Ke Li[1], Thomas Nagel[2]

**AFFILIATIONS**

1 Center for Hypergravity Experimental and Interdisciplinary Research, MOE Key Laboratory of Soft Soils and Geoenvironmental Engineering, Zhejiang Univ., Hangzhou 310058, China.
2 Geotechnical Institute, Technische Universität Bergakademie Freiberg, Freiberg 09599, Germany
* Correspondence and requests for materials should be addressed to Xu W.J. (email: wenjiexu@zju.edu.cn) or Chen Y.M. (email: chenyunmin@zju.edu.cn)

**ABSTRACT**

Scaled physical modeling is an important means to understand the behavior of fluids in nature. However, a common source of errors is conflicting similarity criteria. Here, we present using hypergravity to improve the scaling similarity of gravity-dominated fluid convection, e.g. natural convection and multi-phase flow. We demonstrate the validity of the approach by investigating water-brine buoyant jet experiments conducted under hypergravity created by a centrifuge. Results show that the scaling similarity increases with the gravitational acceleration. In particular, the model best represents the prototype under $N^3g$ with a spatial scale of $1/N$ and a time scale of $1/N^2$ by simultaneously satisfying the Froude and Reynolds criteria. The significance of centrifuge radius and fluid velocity in determining the accuracy of the scaled model is discussed in the light of Coriolis force and turbulence. This study demonstrates a new direction for the physical modeling of fluids subject to gravity with broad application prospects.

**INTRODUCTION**

Physical modeling is a fundamental approach for humans to study fluid [1], especially for turbulence, the partial differential equations describing which have no general analytical solutions [2]. Physical modeling can provide extra boundary conditions or suggest simplifying assumptions for the partial differential equations. The experimental data and observations can furthermore validate analytical or numerical solutions [3]. Physical modeling of fluid flow includes field and scaled experiments. Field experiments are, in principle, reliable and accurate, but the cost can be unacceptable due to the large spatial and temporal scale of the fluid convection. While laboratory scaled models are more convenient with low economical and time costs, controllable boundary conditions and observable output parameters. Thus, scaled physical modeling has been widely accepted and used to study many kinds of fluid flow problems, including combustion and flames [4,5,6], atmospheric currents [7], gas leakage and dispersion [8,9,10], aerodynamics (wind tunnel) [11], dam break [12], sea wave [13] and many more.

Similarity analysis is the key technique to render the scaled models a physically adequate representation of the prototype [14]. Non-dimensional numbers, including the Reynolds number, Froude number, Mach number, Euler number, etc., have been proposed to define the similarity of the scaled model to its prototype. Theoretically, the flow field of a scaled model will be exactly the same as that of its prototype if all the similarity numbers are the same. However, it is nearly impossible to meet all associated similarity criteria simultaneously [15-21]. The main reason is that fundamental physical properties such as density, viscosity, heat transfer coefficient, etc. are usually regarded as equal as long as the same fluids are used, since the simultaneous adjustment of all these physical properties is difficult [17,18]. There are no enough adjustable variables for similarity balance. Thus, most researchers will determine the principal acting force and maintain the corresponding similarity criterion during the scaling of a certain problem while neglecting others. Arini et al. [22] maintained the Froude number in a 1:8 scaled modeling of the smoke movement in a basement, but the Reynolds number was different from that of the real-scale system. Yu [23] proposed a scaling relationship for fire suppression by water sprays, and only Froude number is maintained. Wang et al. [13] maintained the Froude number in the modeling of an aircraft as well.

However, this method is not satisfactory for problems with two main forces. For example, viscosity and gravity are two of the most common dynamics in fluids. They are both relevant in natural convection [24], gravitational or buoyant convection [25] and multi-phase flow [26]. Ignoring one of them in scaled models will certainly result in errors. Especially in transient fluid problems, the errors will accumulate and the final flow phenomena can be totally different. Matching both criteria has been impossible in the past as long as the gravitational acceleration entering the Froude number had to been seen as a constant. The conflict of the two most common similarity criterion can be expressed by,

$$\begin{cases} Re = \rho v l / \mu, Fr = v/\sqrt{gl} \\ \dfrac{l_m}{l_p} = \dfrac{1}{N}, \dfrac{\rho_m/\mu_m}{\rho_p/\mu_p} = 1, \dfrac{g_m}{g_p} = 1 \end{cases} \rightarrow \begin{cases} \dfrac{Fr_m}{Fr_p} = 1 \rightarrow \dfrac{v_m}{v_p} = \sqrt{\dfrac{1}{N}}, \dfrac{Re_m}{Re_p} = \sqrt{\dfrac{1}{N^3}} \\ \dfrac{Re_m}{Re_p} = 1 \rightarrow \dfrac{v_m}{v_p} = N, \dfrac{Fr_m}{Fr_p} = \sqrt{N^3} \end{cases}$$
(1)

where, $Re$ is the Reynolds number, $\rho$ is density, $v$ is velocity, $l$ is the characteristic scale, $\mu$ is the dynamic viscosity of the fluid, $Fr$ is the Froude number, $g$ is the gravitational acceleration, the subscript $m$ and $p$ represent the model and prototype respectively.

However, the "gravity" can be an adjustable variable (gravity = $ng$) in an accelerating system, e.g. in a rotational centrifuge or in a speeding rocket, enabling a breakthrough in the conflict of similarity criteria: The Reynolds number and Froude number can be maintained simultaneously under $N^3g$ (Eq. 2),

$$\dfrac{g_m}{g_p} = n \rightarrow \begin{cases} \dfrac{Fr_m}{Fr_p} = 1 \rightarrow \dfrac{v_m}{v_p} = \sqrt{\dfrac{n}{N}}, \dfrac{Re_m}{Re_p} = \sqrt{\dfrac{n}{N^3}} \\ \dfrac{Re_m}{Re_p} = 1 \rightarrow \dfrac{v_m}{v_p} = N, \dfrac{Fr_m}{Fr_p} = \sqrt{\dfrac{N^3}{n}} \end{cases}$$
(2)

when $n = N^3 \rightarrow \dfrac{Fr_m}{Fr_p} = 1, \dfrac{Re_m}{Re_p} = 1$

In recent years, the capacity and the size of centrifuges has been growing, making large-size scaled models feasible. For example, the largest centrifuge built by the Soviet Union has an effective spinning radius of 5.5 m, 500 g hypergravity, and 3 t loading capacity [27]. A beam centrifuge with a large rotatable blanket has been widely used in physical modeling in the area of soil mechanics [28,29], structural mechanics and tectonic movements. The "hypergravity" results in more realistic stress fields for scaled soil or structural models, and can accelerate seepage phenomena that take years in reality. However, there has been little research about the physical modeling of convective problems in a centrifuge.

In this paper, scaled physical modeling under hypergravity has been proposed to increase the similarity of physical models in convective flow problems. Centrifugal hypergravity experiments of buoyant jets are conducted as a typical example of convective problems subject to gravitational and viscous forces. The gravitational acceleration, initial jet velocities are designed based on the similarity analysis using Reynolds and Froude criteria. Experimental results are compared to discuss the feasibility and applicability of the hypergravity experiment. In the end, the further application possibilities of the hypergravity model is discussed to more flow phenomena subject to gravitational and other forces.

## METHODS

A buoyant jet is a typical mixed convective flow problem [30], encountered, for example, in waste water disposal [31], gas leakage [9,10], flame jets [8], geothermal systems [32], solution mining [33,34] and glacial melting [35]. The gravitational force is a dominant force in a buoyant jet due to the density difference between the jet fluid and the ambient fluid. The viscous force is considerable as well, due to the interaction between the jet fluid and the entrained ambient fluid. In this paper, brine-water buoyant jet experiments are conducted as a typical example to discuss the scaling similarity of fluid convection. The experiments have been conducted in a specially designed water tank, as shown in Fig. 1a. Prototype experiments (normal gravity) and 1/$N$ scaled models under normal/hyper gravity have been conducted for similarity comparison. The "hypergravity" was produced by a rotational beam centrifuge (Fig. 1b), ZJU-400 in Zhejiang University.

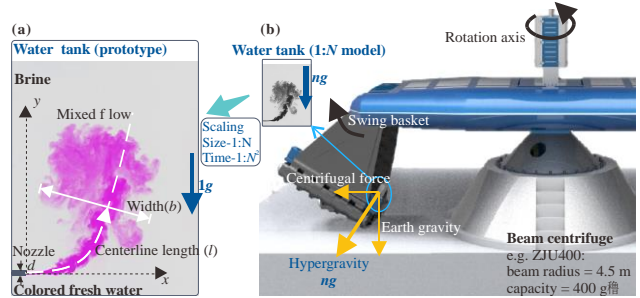

**Figure 1. Sketch of the buoyant jet experiment. (a)** Colored fresh water (1.0 g/cm$^3$) is injected into a water tank filled with brine (1.3 g/cm$^3$). **(b)** A reduced scale model is placed in the basket of a beam centrifuge, the rotation of which will produce the required "hypergravity". The mixing process is recorded by a camera and then analyzed through image processing.

## RESULTS AND DISSCUSSIONS

### Section 1, similarity of hyper-g and normal-g models

12 groups of buoyant jet experiments have been conducted with different reduced scales $N$ ($N$=2, 3, 4) and different initial jet velocities. In each group, prototype, hyper-g model and normal-g model are set (Fig. 2a). The parameters are set according to Eq.1 and Eq.2, only Froude number is maintained in normal-g model, while both Froude and Reynolds number are maintained in hyper-g model ($N^3$g). Detailed parameters are provided in Table A1.

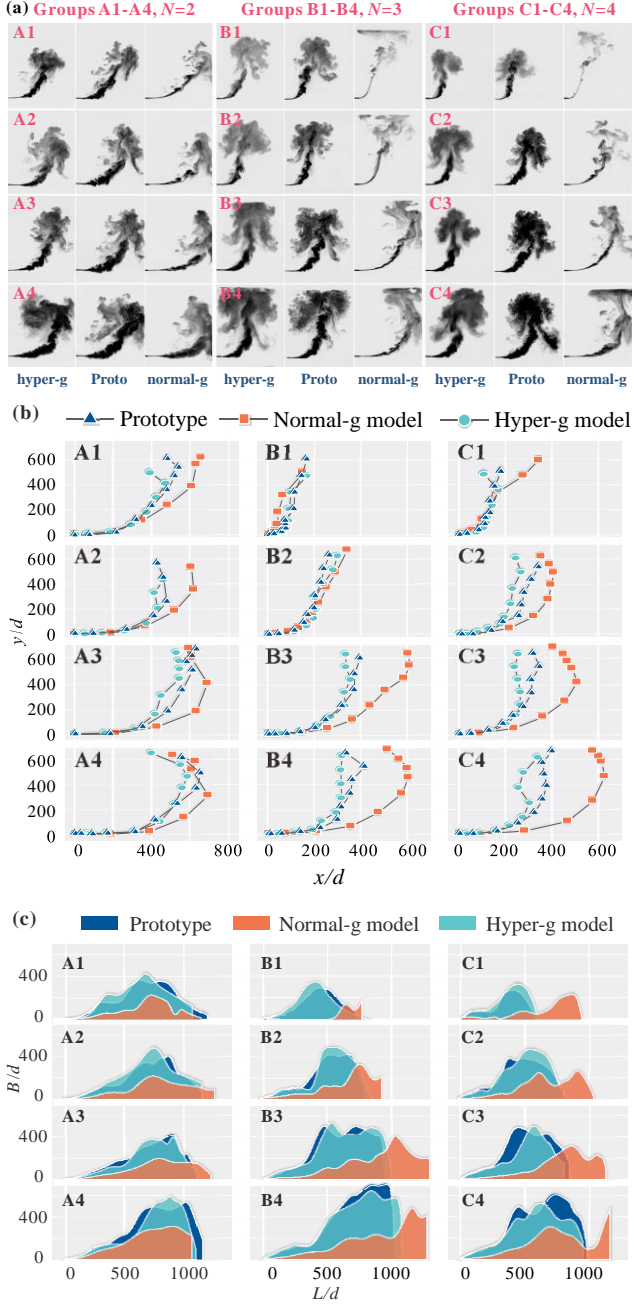

**Figure 2.** Grayscale images of the mixed flow. (a) 12 groups of the hyper-g 1/N model, normal-g 1/N model and prototype. The size of the prototype has been normalized to facilitate the comparison. (b) Normalized coordinates of the centerline of the mixed flow. (c) Normalized width of the mixed flow along its centerline. The centerline and width of the hyper-g model (light blue) is much closer to that of the prototype (dark blue) compared with the normal-g model (red). The advantage is more noticeable when the reduced scale is large and the initial jet velocity is large.

It turns out that the shapes of the mixed flow of the hyper-g model are quite close to those of the prototype, while those of the normal-g model are quite different (a quantitatively comparison is shown in Fig. 2b and Fig. 2c). The Froude numbers are the same in the hyper-g model, normal-g model and prototype, which means that the effect of gravity is the same in all three [36]. In the normal-g model, the Reynolds number is smaller than that of the prototype, which means that the mixing is less intense and the momentum loss is slower. Thus, the horizontal projection distance of the mixed flow is longer than that of the prototype. And the width change is far slower than that of the prototype due to the lower Reynolds number and less viscous entrainment. In the hyper-g model, the centerline is much closer to that of the prototype since they have the same Reynolds numbers and Froude numbers. The width change and the turbulent transition are almost identical with those of the prototype, especially in the first half, since the Reynolds criteria is honored. To further discuss the temporal scaling effect of the hyper-g model, the shapes of the mixed flow of the prototype and of the hyper-g model are compared in Figure.3 (only A1 group is shown to save space, the spatial scale is 1:2). A 1:4 (1: $N^2$) temporal scale is observed, the mixed flow of the hyper-g model at time $t$ are very similar to that of the prototype at time $N^2t$. Even the development of the turbulence are almost the same. In summary, the hyper-g scaled model of buoyant jet has much more similarity with the prototype compared with the normal-g model. A $N^3$g model can duplicate the prototype with a spatial scale of $1/N$ and a time scale of $1/N^2$.

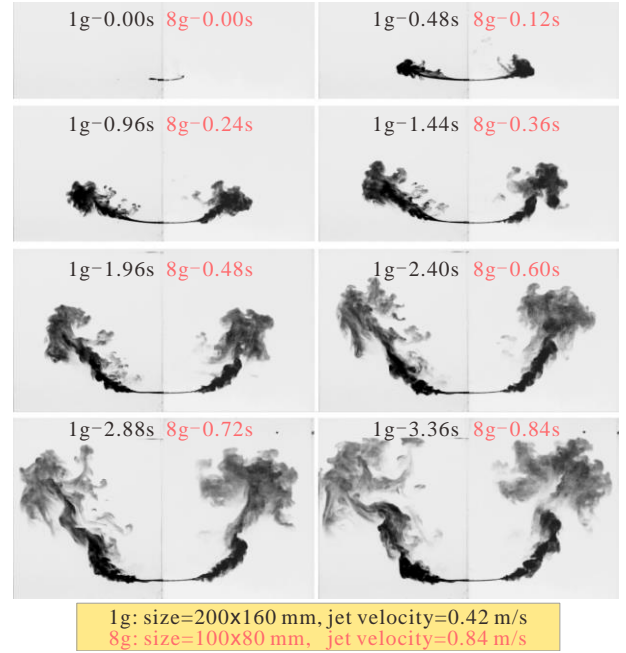

**Figure 3** Comparison of the hyper-g 1/$N$ model at time t and the prototype at $N^2$t. The hyper-g model is similar to the prototype, the spatial scale is 1:$N$, and the temporal scale is 1:$N^2$.

## Section 2, similarity of models with different gravity levels

Working with a reduced hypergravity loading compared to the full $N^3$ scaling can be relevant when considering physical systems (prototypes) which are very large and require a high geometrical scaling factor $N$. The applicability of the hypergravity model is further discussed. Four groups (D, E, F, G) of buoyant jet experiments are conducted. Each group contains a prototype and five 1/4 scaled models with different gravitational acceleration (1g, 8g, 10g, 27g, 64g). The initial jet velocities of the prototypes are different in the 4 groups. To maintain the Froude number, the jet velocity of the models is $(n/4)^{0.5}$ of that of the prototype (Eq. 2). The Reynolds number is different in the models. Detailed parameters can be found in Table A2.

The images of the mix flow of the four groups are shown in Fig. 4a. Along with the increase of the gravitational acceleration, the horizontal projection distance decreases and gets closer to that of the prototype. The reason is that increasing gravity restores the importance of buoyancy as a driving force in relation to the inertia imparted to the jet by its initial velocity. In addition, the increase of gravity allows a higher velocity when maintaining the Froude number, thus a higher Reynolds number (Eq. 2). The viscous effect gets larger and the mixing is more intense, the mixed flow width increases faster and get closer to that of the prototype.

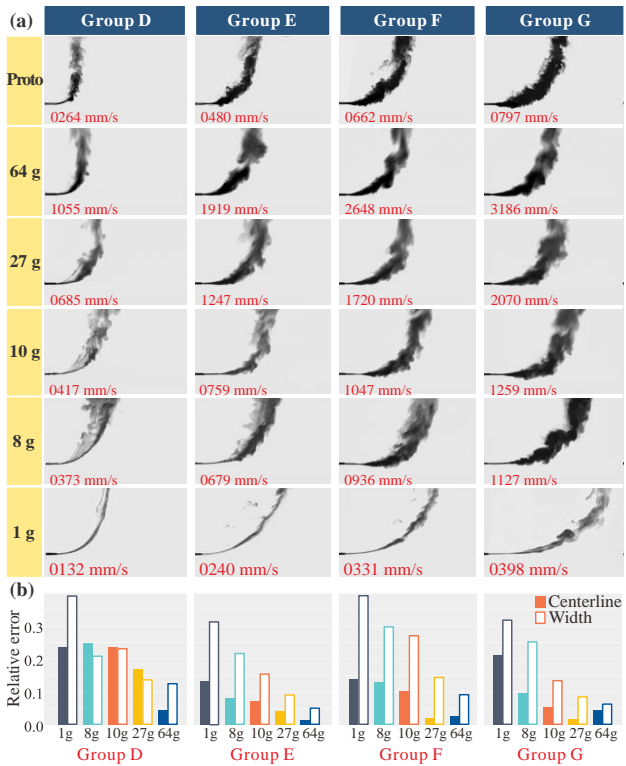

**Figure 4** (a) Grayscale image of the mixed flow of groups D, E, F and G. (b) Relative errors of the centerline positions and widths of the models (1g – 64g). The errors decrease with the increase of g-level in all the four groups. But the improvement is not obvious with higher g-level and higher initial jet velocity, due to the presence of Coriolis force and the fully developed turbulence.

The relative error coefficients are calculated based on the average errors between the centerline/width of the models and those of the normalized prototype, cf. Fig. 4b. The errors decrease with the increase of hypergravity. The average relative error coefficients of the centerline position of the four groups are 0.179 (1 g), 0.139 (8 g), 0.115 (10 g), 0.062 (27 g), and 0.032 (64 g). The average relative error coefficients of the width of the four groups are 0.357 (1 g), 0.243 (8 g), 0.196 (10 g), 0.113 (27 g), and 0.081 (64 g). The results demonstrate well that hypergravity can improve the similarity of the scaled experiments even when the g-level is less than $N^3$.

However, there are unexpected deviations in group F and group G. The horizontal projection distance of the 64g-model gets shorter than that of the prototype, and the errors get larger than the 27g-model (Fig. 4b). It cannot be explained by the similarity theory. That probably occurs due to the presence of the Coriolis acceleration caused by the rotational centrifuge. The Coriolis acceleration is related to the relative velocity of the mass point, and the angular velocity of the rotation. Thus, this deviation is larger in group G with higher velocity compared with group F.

$$a_{co} = 2\omega v \qquad (3)$$

When the initial jet velocity is low, the error caused by the Coriolis acceleration can be neglected (group D, E). But when the initial jet velocity is high, the error gets noticeable. To reduce this error, high-velocity problems should be avoided for scaled modeling in centrifugal hyper-gravity system (this problem does not exist in other hyper-gravity system, e.g. in a linear accelerating plane or on other planets). On the other hand, the radius of the centrifuge can be increased to reduce the required angular velocity $\omega$ (for producing designed hypergravity), and thus the Coriolis acceleration will be reduced according to Eq. 3. Preferably, the Coriolis force can be included as a momentum source term and applied in numerical calculations [37]. By combining the numerical calculation and the experimental results, the error can be corrected.

In addition, it is worth mentioning that when the gravitational acceleration is more than doubled from 27g to 64g, while the difference between the model and the prototype decreases only slightly, especially in high-speed cases. The reason is that the effect of the Reynolds number on the flow field properties decreases when it is larger than the critical Reynolds number (about 2000 for a round jet [38,39]). If this error is allowed, the Reynolds criterion can be relaxed as,

$$\begin{cases} Re_m \geq Re_{cr} \, (Re_p \geq Re_{cr}) \\ Re_m < Re_{cr} \, (Re_p < Re_{cr}) \end{cases} \qquad (4)$$

where, $Re_m$ is the Reynolds number of the model, $Re_{cr}$ is the critical Reynolds number, $Re_p$ is the Reynolds number of the prototype.

Thus, the requirements of hypergravity for a 1: $N$ scaled model can be written as,

$$\begin{cases} n = N^3 \ (Re_p \leq Re_{cr}) \\ n = \dfrac{Re_{cr}}{Re_p} N^3 \ (\sqrt{\dfrac{1}{N^3}} Re_p \leq Re_{cr} < Re_p) \\ n = 1 \ (\sqrt{\dfrac{1}{N^3}} Re_p \geq Re_{cr}) \end{cases} \quad (5)$$

where, n is the ratio of the hypergravity to the earth's gravity, $N$ is the reduced scale of the model.

It can be seen from Eq. 5 that the hypergravity model is suitable for convective problems with low Reynolds number. If $\sqrt{\dfrac{1}{N^3}} Re_p \geq Re_{cr}$, normal gravity scaled modeling is adequate, there is no need for hypergravity.

## CONCLUSIONS

Hypergravity experiments have been performed to ensure that scaled physical modeling of fluid convection under the influence of gravitational forces remains representative of the physical prototype system by honoring the relevant dimensionless similarity criteria. To the best of the authors' knowledge this is the first physical modeling study of convective flow problems that duplicate the prototype with a spatial scale of $1/N$ and a time scale of $1/N^2$, matching both Froude and Reynolds criteria, by using experimental hypergravity facilities. It demonstrates a new method for the lab-scale investigation of fluid convection under the influence of gravitational forces. Considering the wide application of physical modeling of fluids in areas of physics, thermodynamics, meteorology, aerodynamics, hydraulic engineering, energy and materials science, environmental science etc., the proposed hypergravity method will have a broad application prospect. Here, we summarize and discuss the principal results.

1) It is derived that the Reynolds number of a normal scaled model is far smaller than that of the prototype when maintaining the Froude number. This difference is reduced by introducing a variable gravitational acceleration in the hypergravity experiments. The Reynolds number and the Froude number of a 1: $N$ scaled model under $N^3$ g will both match those of the prototype exactly.

2) Water-brine buoyant jet experiments have been conducted to compare the similarity of the hypergravity and normal-gravity models. Results show that the hyper-g model under $N^3$g (created by a centrifuge) is similar to the prototype, while the normal-g model is not. The spatial scale of the hyper-g model is 1:$N$ and the time scale is 1:$N^2$.

3) To investigate the similarity of the scaled model with different gravitational accelerations, water-brine buoyant jet experiments have been conducted, including prototype and 1:4 scaled models under 1 g, 8 g, 10 g, 27 g and 64 g. The results show that the similarity of the models increase with the gravitational acceleration. The hypergravity level of $N^3$g may not be achievable when considering large physical systems (prototypes) requiring a high geometrical scaling factor $N$. But as shown here, $n>1$ may still provide acceptable results while $n=1$ may not.

4) A limitation of using a rotational acceleration system is the presence of Coriolis effects. In the present case, the Coriolis acceleration causes a deviation when the initial jet velocity is high. Increasing the radius of the centrifuge, and introducing numerical simulations will help to reduce or assess the error caused by the Coriolis acceleration.

5) Increasing the gravity has a limited effect on the similarity when turbulence is fully developed (Reynolds number is higher than the critical Reynolds number). The required hypergravity for the similarity of the model is derived. It shows that the hypergravity model is more relevant for convective problems with low Reynolds numbers.

**Outlook**

It was demonstrated that using hypergravity facilities allows experimental designs based on an extra variable in order to maintain the similarity between a reduced-scale model of fluid convection with gravitational and viscous forces and the original physical system. In addition, the proposed method could be used in fluid mechanical problems containing gravitational and other forces, e.g. surface tension. The hypergravity modeling of the capillary effects and multiphase flow, which is a key problem in energy exploration, metallurgy, fuel cell design, the chemical industry and many other fields, will be among the next problems studied. The surface tension acts in addition to gravity in these problems, and the variable gravitational acceleration might help to resolve the conflict between the similarity criteria of the two forces. The surface tension and gravity similarity criteria can be both satisfied as derived in Eq. 6. Thus, it can be said that the hypergravity approach will have a general applicability to the modeling of more fluid-mechanical phenomena in gravitational field.

$$\begin{cases} \dfrac{\rho_m / \mu_m}{\rho_p / \mu_p} = 1, \dfrac{l_m}{l_p} = \dfrac{1}{N}, \dfrac{g_m}{g_p} = N^2, \dfrac{v_m}{v_p} = \sqrt{N} \\ \dfrac{Fr_m}{Fr_p} = 1, \dfrac{W_m}{W_p} = 1 \end{cases} \quad (6)$$

where, $W_m$ and $W_p$ are the Weber number of the model and of the prototype.

## ACKNOWLEDGEMENTS

The research was supported by Basic Science Center Program for Multiphase Evolution in Hypergravity of the National Natural Science Foundation of China (No. 51988101), China

Postdoctoral Science Foundation (Grant No. 2018M642433).

## DATA AVAILABILITY

The data that support the findings of this study are available from the corresponding author upon reasonable request.

# APPENDEX

Table A1 Parameters of groups A, B, C.

| | Parameters | | Prototype | Hyper-g | Normal-g |
|---|---|---|---|---|---|
| **Group A** <br> **N=2** | Water tank/mm | | 200×160×6 | 100×80×3 | 100×80×3 |
| | Jet mouth/mm | | 2 | 1 | 1 |
| | Gravity/g | | 1 | 8 | 1 |
| | Velocity <br> mm/s | **A1** | 424.4 | 848.8 | 307.2 |
| | | **A2** | 530.5 | 1061 | 364.5 |
| | | **A3** | 636.6 | 1273.2 | 460 |
| | | **A4** | 848.8 | 1697.6 | 593.6 |
| **Group B** <br> **N=3** | Water tank/mm | | 300×240×9 | 100×80×3 | 100×80×3 |
| | Jet mouth/mm | | 3 | 1 | 1 |
| | Gravity/g | | 1 | 27 | 1 |
| | Velocity <br> mm/s | **B1** | 262.6 | 786.2 | 154.4 |
| | | **B2** | 351.7 | 1055.2 | 211.7 |
| | | **B3** | 531.6 | 1593.1 | 307.2 |
| | | **B4** | 620.7 | 2131.1 | 364.5 |
| **Group C** <br> **N=4** | Water tank/mm | | 400×320×12 | 100×80×3 | 100×80×3 |
| | Jet mouth/mm | | 4 | 1 | 1 |
| | Gravity/g | | 1 | 64 | 1 |
| | Velocity <br> mm/s | **C1** | 262.6 | 1055.2 | 135.3 |
| | | **C2** | 531.6 | 2129.5 | 269 |
| | | **C3** | 662.1 | 2648.3 | 326.3 |
| | | **C4** | 795.8 | 3186.3 | 402.7 |

Table A2 Parameters of groups D, E, F, G.

| Name | Water tank/mm | Jet mouth/mm | Gravity /g | Velocity/mm/s | | | |
|---|---|---|---|---|---|---|---|
| | | | | Group D | Group E | Group F | Group G |
| **Prototype** | 400×320×12 | 4 | 1 | 264 | 480 | 662 | 797 |
| **Model** | 100×80×3 | 1 | 64 | 1055 | 1919 | 2628 | 3186 |
| | 100×80×3 | 1 | 27 | 685 | 1247 | 1720 | 2070 |
| | 100×80×3 | 1 | 10 | 417 | 759 | 1047 | 1259 |
| | 100×80×3 | 1 | 8 | 373 | 679 | 936 | 1127 |
| | 100×80×3 | 1 | 1 | 132 | 240 | 331 | 398 |